\documentclass[twocolumn]{aastex7}

\usepackage{CJK}
\usepackage{amsmath}
\usepackage{graphicx}

  \renewcommand{\d}{\mathrm d}
    \newcommand{\vct}[1]{\boldsymbol{#1}}

  \newcommand{\Sell}{\overline{\sf S}}
  \newcommand{\avg}  [1]  { \langle #1 \rangle }

\begin{document}
\begin{CJK*}{UTF8}{gbsn}

\title{Estimation of Effective Viscosity to Quantify Collisional Behavior in Collisionless Plasma}

\author[orcid=0000-0003-2160-7066]{Subash Adhikari}
\email{subash@udel.edu}
\affiliation{Department of Physics and Astronomy,
University of Delaware, Newark, DE 19716, USA}

\author[orcid=0000-0001-7063-2511]{Carlos Gonzalez}
\email{carlos.gonzalez1@austin.utexas.edu}
\affiliation{Department of Physics, The University of Texas at Austin, Austin, TX 78712, USA}

\author[orcid=0000-0003-2965-7906]{Yan Yang (杨艳)}
\email{yanyang@udel.edu}
\affiliation{Department of Physics and Astronomy,
University of Delaware, Newark, DE 19716, USA}

\author[orcid=0000-0002-2814-7288]{Sean Oughton}
\affiliation{Department of Mathematics, University of Waikato, Hamilton 3240, New Zealand}
\email{sean.oughton@waikato.ac.nz}

\author[orcid=0000-0003-4168-590X]{Francesco Pecora}
\email{fpecora@udel.edu}
\affiliation{Department of Physics and Astronomy,
University of Delaware, Newark, DE 19716, USA}

\author[orcid=0000-0002-6962-0959]{Riddhi Bandyopadhyay}
\email{riddhib@princeton.edu}
\affiliation{Department of Astrophysical Sciences, Princeton, NJ 08544, USA}

\author[orcid=0000-0001-7224-6024]{William H. Matthaeus}
\email{whm@udel.edu}
\affiliation{Department of Physics and Astronomy,
University of Delaware, Newark, DE 19716, USA}

\correspondingauthor{Subash Adhikari}
\email{subash@udel.edu}

\begin{abstract}
While dissipation in collisional plasma is defined in terms of viscosity and resistivity, the exact functional form of dissipation 
i.e., the so-called \emph{dissipation function}  in 
nearly 
collisionless plasma is unknown. Nevertheless, previous studies have suggested 
that there exists viscous-like energy conversion in collisionless plasma with scaling characteristics analogous to collisional plasma, 
and in particular that the average dissipation
is proportional to the square of the rate of strain as in hydrodynamics. 
In this study, using $2.5$D 
kinetic particle-in-cell (PIC) simulation of collisionless plasma turbulence, we provide an estimate of effective viscosity 
at each scale, 
obtained via a scale-filtering approach. 
We then compare the turbulent dynamics of the PIC simulation with that from MHD and two-fluid simulations in which with the viscosity is equal to the effective viscosity estimate obtained from the PIC simulation. 
We find that the global behavior in these MHD and two-fluid simulations has a striking similarity with that in its kinetic/PIC counterpart. 
In addition, we explore the scale dependence of the effective viscosity, and discuss implications of this approach for space plasmas. 
\end{abstract}

\section{Introduction}
\end{CJK*}
\linenumbers

A fundamental assumption regarding dissipation in collisional plasmas is that Coulomb collisions are strong enough to 
drive the system to a 
local equilibrium. However, in many (nearly) collisionless plasmas this is not true, since such plasmas have very long collisional mean free paths, and are therefore not able to 
  `quickly' 
establish local equilibrium.
As a result, the standard closures often employed in the collisional case, that relate energy dissipation to viscosity and resistivity, 
are inapplicable to 
  nearly 
collisionless plasmas~\citep{braginskii1965transport}. 

Intriguingly, however, power spectra obtained from
   nearly
collisionless plasma systems, such as the solar wind or appropriate numerical simulations, 
display behavior similar to that of hydrodynamic (i.e., strongly collisional) turbulence. 
For example, the wavenumber spectra often display an inertial range with a Kolmogorov-like $k^{-5/3}$ scaling, followed by a sharp transition to steeper spectra at larger wavenumbers,
suggesting that the ``collisionless'' plasma may not be perfectly collisionless in a 
    strict 
sense. 
In addition, recent advances connected with simulations and high-resolution spacecraft observations have suggested that pressure-strain interaction can describe the energy transfer at smaller kinetic scales and therefore represents the dissipation 
function in collisionless plasmas processes, 
such as turbulence~\citep{yang2022pressure} and reconnection~\citep{adhikari2024scale}. 
These features illustrate that 
   (nearly)
collisionless plasma can 
     in fact 
behave in ways that are analogous to collisional dynamics, and that there might be ways to quantify the collisional-like effects, even potentially achieving the goal of determining a collisionlees closure that captures dissipation in the 
case of collisionless turbulence~\citep{matthaeus2020pathways, pezzi2021dissipation}.

Traditionally it has not been uncommon to find the electromagnetic work ${\vct{j} \cdot \vct{E}}$
identified as a measure of dissipation --a practice particularly familiar in the reconnection community
\citep{zenitani2011new}.  However examination of the Vlasov--Maxwell system \citep{yang2017energy,yang2019scale}
reveals that while the rate of conversion of electromagnetic energy into kinetic energy is indeed determined by 
$\vct{j} \cdot \vct{E}$, the species-dependent pressure-strain interactions provides the channels for conversion of 
flow energy into the internal energy of the respective species.  Both electromagnetic work and pressure-strain are integral to understanding the cascade and pathways to dissipation, and research has progressed on both. 
Several recent studies \citep{bandyopadhyay2023collisional,yang2024effective}  have explicitly shown that the \emph{average}
of the electromagnetic work 
  $ \vct{j} \cdot \vct{E} $ \citep{zenitani2011new} 
scales as the square of the electric current density, $\vct{j}^2$, when conditioned over 
 $|\vct{j}|$.
Similarly, the average of one of the ingredient of pressure-strain interaction
(also called pressure work), 
when conditioned on a threshold of $D$,
has been shown to scale as the (trace of the) squared 
velocity strain rate tensor, 
   $D^2 = {\sf S}_{ij} {\sf S}_{ji}$.
     (see \S\ref{sec:theory} for definitions). 
These results suggest the existence of collisional-\emph{like} dissipation in so-called collisionless plasmas. Moreover, they enable quantification of effective viscosity and resistivity in collisionless plasma, which may then be used to help provide relevant estimates of energy dissipation.
However, neither of these results involving \emph{conditional averages} 
of the dissipation provide pointwise relationships between local dissipation and the respective 
dissipation coefficients. Likewise the conditional averages provide no information about scale dependence of the 
transport coefficients. 
Both of these deficiencies
are resolved for the viscous transport in the  
present alternative strategy. In this study, we consider a different approach to obtain estimates for effective ion viscosity---based on a scale-filtering technique---by examining the evolution of bulk flow energy using two different plasma models: Magnetohydrodynamics (MHD) and Vlasov--Maxwell (VM). 
By comparing the different terms in their respective energy evolution equations, we obtain an analogy between them that leads to an estimate for an effective viscosity.

    \section{Theory}\label{sec:theory}

Before diving into the specifics of this study, we provide a brief overview of the scale-filtering technique. 
Scale-filtering \citep{germano1992turbulence} is based on a properly defined filtering kernel 
  $ G_\ell = \ell^{-d} G( \vct{r}/\ell ) $
which, when convolved with any field 
  $f(\vct{x}, t)$, 
only maintains information about $f$ at length scales 
larger than the filtering scale $\ell$. 
Here $G(\vct{r})$ is a  normalized window function, usually chosen to be spatially compact, that satisfies $\int \d^d r \, G(\vct{r})=1$, where $d$ is the number of dimensions of the system.
For this study, we employ a non-negative boxcar (aka top hat) function for $G$.
The scale-filtered version of $f$, denoted $\overline{f}_\ell(\vct{x},t)$ is defined as the convolution
\begin{equation}\label{eqn:scalefilter}
     \overline{f}_\ell(\vct{x},t) 
     = 
       \int \d^d r \,
           G_\ell( \vct{r} )
                f( \vct{x} + \vct{r}, t ).
\end{equation}
In situations where it is clear from context that the filter-scale is $\ell$, 
we will often use 
    $\overline{f}$ 
to mean 
    $\overline{f}_\ell $.

Likewise, the density-weighted scale-filtering of $f(\vct{x},t)$ is defined as (note the over\emph{tilde})
\begin{equation}\label{eqn:favrefilter}
     \widetilde{f}_\ell(\vct{x},t) 
     = 
     \frac{\overline{\left[\rho(\vct{x},t) f(\vct{x},t)\right]}_\ell}{\overline{\rho}_\ell(\vct{x},t)},
\end{equation}
where $\rho(\vct{x}, t)$ is the 
  mass
density. This is also called the Favre-filtered field~\citep{favre1969statistical,aluie2013scale}.

Using these techniques one can obtain evolution equations for the scale-filtered kinetic energy associated with MHD and VM models.
These equations, and in particular their terms related to dissipation, 
will form the basis of the analysis below. 

     \subsection{Incompressible MHD formalism}
In incompressible MHD, the time evolution of the coarse-grained kinetic energy density~\citep{aluie2017coarse} at any scale $\ell$ 
   obeys
\begin{equation}
\label{eqn:KE_MHD}
    \partial_t E^{\text{MHD}}_f + \nabla \cdot \vct{J}_{u}^{ \text{MHD}} = -\Pi^{\text{MHD}}_u - \Lambda_{ub}^{\text{MHD}} - 2\mu \Sell \mathrm{:} \Sell.
\end{equation}
Here, 
 $ E^{\text{MHD}}_f
    = \rho \frac{\lvert \overline{\vct{u}} \rvert^2}{2}$ 
is the filtered
kinetic energy density per unit volume
associated with the  coarse grained bulk flow velocity $\overline{\vct{u}}$,
where mass density of the fluid is $\rho$. 
Likewise, 
$ \vct{J}_{u}^{\text{MHD}}
 =
 (\rho \frac{\lvert \overline{\vct{u}} \rvert^2}{2} + \overline{P}) \overline{\vct{u}} 
 + \overline{\tau} \cdot \overline{\vct{u}} 
 - (\overline{\vct{u}}\cdot \overline{\vct{b}}) \overline{\vct{b}} 
 - \mu \nabla (\frac{\lvert \overline{\vct{u}} \rvert^2}{2})$ 
is related to various contributions to transport of 
filtered kinetic energy. 
This includes the advective flux of large-scale kinetic energy, 
the influence of the Poynting flux, involving 
the coarse grained magnetic field $\overline{\vct{b}}$, the filtered total
  scalar
pressure $\overline{P}$, 
 the sub-filterscale nonlinear stresses
 $\overline{\tau}$, defined as the sum of the subscale Reynolds stress $\overline{\tau}^u=\rho\left[\overline{\vct{u}\vct{u}} - \overline{\vct{u}}\,\overline{\vct{u}} \right]$ and the subscale Maxwell stress $\overline{\tau}^b=-\rho\left[\overline{\vct{b}\vct{b}} - \overline{\vct{b}}\,\overline{\vct{b}} \right]$. 
$\overline{\tau}$ characterizes the force acting on scales larger than $\ell$ due to fluctuations occurring at scales less than $\ell$, 
and $\mu$ is the dynamic viscosity. 
Similarly, the first term on the right-hand side (RHS), 
$\Pi^{\text{MHD}}_{u}=\Sell \mathrm{:}\overline{\tau} $, is the subgrid scale 
kinetic energy flux that quantifies the transfer of kinetic energy across scales $\ell$, where $\sf S$ is the rate-of-strain tensor
 with  
 $ \Sell = (\nabla \overline{\vct{u}} + \nabla \overline{\vct{u}}^T)/2$
 its scale-filtered version. 
The second term $\Lambda_{ub}^{\text{MHD}}
 =
 \overline{\vct{b}} \cdot \Sell \cdot \overline{\vct{b}}$ 
represents the effect of the large-scale flow to stretch or bend the magnetic field lines, 
and the final term on the RHS,
$2\mu \Sell \mathrm{:} \Sell$ is the dissipation of (filtered) kinetic energy due to viscous effects. 
For a detailed description of the terms in Eq.~\eqref{eqn:KE_MHD}
see
\citet{aluie2017coarse}.

  \subsection{Vlasov--Maxwell Formalism}
One can also obtain 
the analog of Eq.~\eqref{eqn:KE_MHD} 
for the VM system
  \citep{matthaeus2020pathways, yang2022pressure}.
The evolution of the filtered kinetic energy of each species $\alpha$ at a given scale $\ell$ follows
\begin{equation}
\label{eqn:KE_VM}
    \partial_t E_{f_\alpha}^{\text{VM}} + \nabla \cdot \vct{J}_{u_{\alpha}}^{\text{VM}} = -\Pi_\alpha^{\text{VM}} -\Lambda_{ub_\alpha}^{\text{VM}}-\Phi_\alpha^{uT},
\end{equation}
where 
  (with no sum over $\alpha$ implied) 
$E^{VM}_{f_\alpha} = \frac{1}{2}\overline{\rho}_\alpha \widetilde{\vct{u}}_\alpha^2$ 
is the filtered bulk flow energy for each species $\alpha$; 
$\vct{J}_{u_{\alpha}}^{\text{VM}}
 = 
 \widetilde{E}_{f_\alpha}\widetilde{\vct{u}}_\alpha + \overline{\rho}_\alpha \widetilde{\tau}^u_\alpha\cdot \widetilde{\vct{u}}_\alpha + 
 \overline{{\sf P}}_\alpha \cdot \widetilde{\vct{u}}_\alpha$
is the spatial transport current,
with
 $\overline{{\sf P}}_\alpha $
the pressure tensor,
and
 $ \Pi_\alpha^{\text{VM}} 
   = 
   -(\overline{\rho}_\alpha \widetilde{\tau}^u_\alpha\cdot \nabla)\cdot \widetilde{\vct{u}}_\alpha 
   - \frac{q_\alpha}{c} \overline{n}_\alpha\widetilde{\epsilon}^b_\alpha \cdot \widetilde{\vct{u}}_\alpha$
is the subscale flux of bulk flow energy across scales due to nonlinearities, with $q$ is the charge, $\overline{n}$ is the filtered number density and $c$ is the speed of light.
Here
 $ \widetilde{\tau}^u_\alpha
   =
   \widetilde{\vct{u}_\alpha \vct{u}_\alpha}-\widetilde{\vct{u}}_\alpha \widetilde{\vct{u}}_\alpha$ is the subscale Reynolds stress
and
 $ \widetilde{\epsilon}^b_\alpha
   =
   \widetilde{\vct{u}_\alpha\times \vct{b}}-\widetilde{\vct{u}}_\alpha\times \widetilde{\vct{b}}$ is the subscale electromotive force related to the electric field generated by the subscale magnetic field and subscale velocity. 
Similarly, $\Lambda_{ub_\alpha}^{VM}=-q_\alpha \overline{n}_\alpha \widetilde{\vct{E}}\cdot \tilde{\vct{u}}_\alpha $, with $\tilde{\vct{E}}$ being the coarse grained electric field,
is the rate of conversion
of fluid flow energy into electromagnetic energy through
filtered $\vct{j}_\alpha \cdot \vct{E}$, 
and 
$ \Phi_\alpha^{uT} 
  = 
  - (\overline{{\sf P}}_\alpha \cdot \nabla)\cdot \widetilde{\vct{u}}_\alpha $ 
is the filtered pressure-strain interaction that corresponds to the rate of conversion of flow energy into internal energy for each species $\alpha$. 

 \subsection{Comparing dissipation in the two models}
\begin{figure}
 \centering
  \includegraphics[width=1\linewidth]{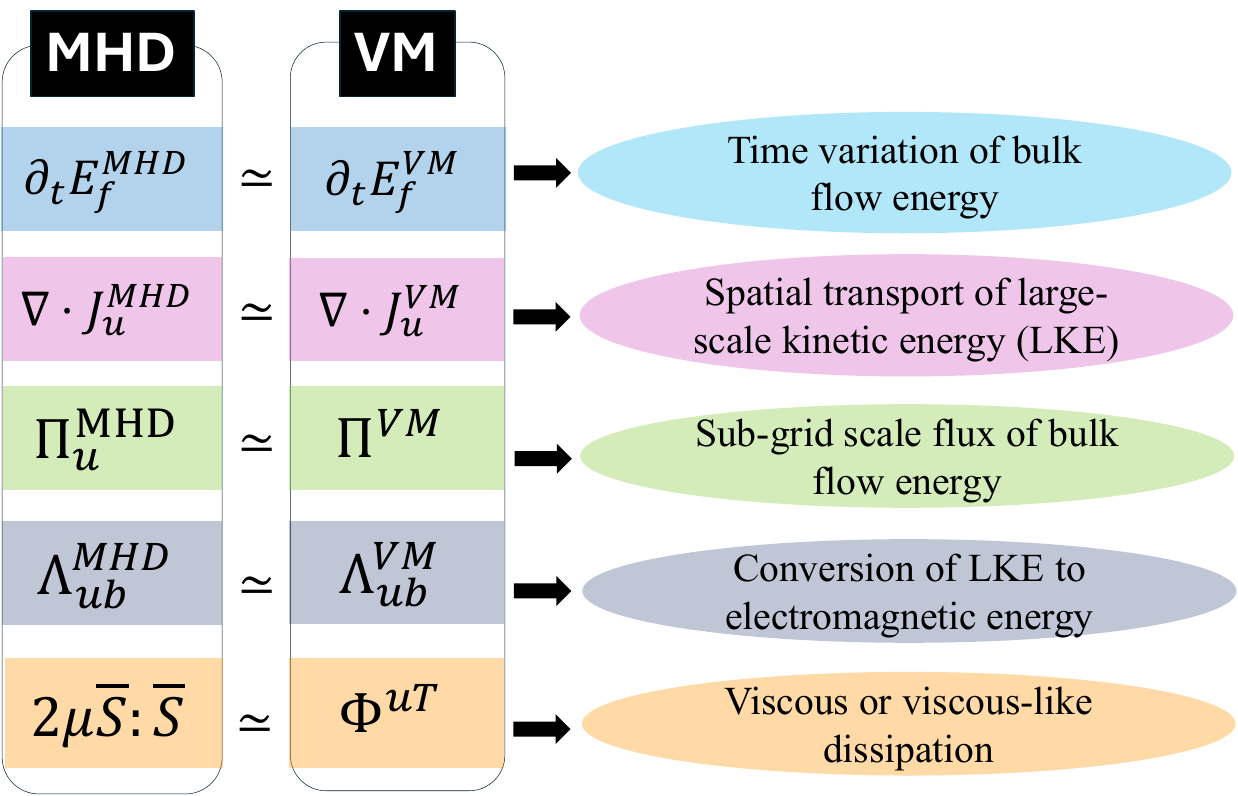}
  \caption{\label{fig:comparing_figure}Comparing different terms in the scale-filtered kinetic energy density equation obtained using the
   Magnetohydrodynamic 
    (MHD) and 
   Vlasov--Maxwell 
  (VM) formalisms.}   
\end{figure}

Even though these two 
   plasma models
are based on different assumptions and approximations, the MHD energy equation should still be comparable to the VM energy equation, 
provided we consider that the momentum and flow energy are carried mainly by the ion species. 
(Consideration of electric current density would likewise involve the electron flow
properties, but we do not consider these here).
Consequently, 
we may make an analogy between corresponding terms in
 Eqns.~\eqref{eqn:KE_MHD} and~\eqref{eqn:KE_VM}; 
    see Fig.~\ref{fig:comparing_figure}. 
Clearly, the first term on the left hand side (LHS) of each equation represents the time rate of change of the flow energy and the second LHS terms represent spatial transport effects.
Likewise, the first term on the RHS represents the 
  subscale flux  (or the kinetic energy cascade) 
term. 
The second RHS term quantifies the conversion of kinetic energy to magnetic energy at scales $> \ell$.

The final terms are also expected to be analogous.
This implies that the (collisionless) $\Phi^{uT}$ term has the same effect as the (collisional) viscous dissipation term $2\mu \Sell \mathrm{:} \Sell$, suggesting that the filtered pressure-strain interaction acts as a viscous dissipation term in 
VM systems.
Recall that most of the momentum is carried by ions and the dissipation term in the ion equation 
would then most closely correspond to the viscous dissipation in the simple single fluid MHD model.~\footnote{One should keep in mind, 
  however, that the MHD model is characterized by a single fluid flow velocity, while in the 
  VM system there are separate velocities associated with each particle species.
  But the disparity of ion and electron masses causes most of the momentum at the scales of current interest to be carried by ions. 
  Very small systems might act differently, as in, for example electron-only reconnection~\citep{phan2018electron}.} 
Thus, using this analogy, we propose to write
\begin{equation}\label{eqn:viscosity_pre}
    2\mu\Sell \mathrm{:} \Sell = \Phi^{uT} = -\overline{p}\nabla \cdot \widetilde{\vct{u}} -\overline{{\Pi}} \mathrm{:} \widetilde{\vct{D}}.
\end{equation}
%
For simplicity, we have removed the subscript $\alpha$ from Eq.~\eqref{eqn:viscosity_pre}, since hereafter
we consider only properties of the ions in the VM case. 
In the usual way, we decompose 
 $ \Phi^{uT} $
using
 $ \overline{p}_\alpha = \overline{{\sf P}}_{\alpha,ii} /3$, 
 $ \overline{\Pi}_{\alpha,ij} 
    =   \overline{{\sf P}}_{\alpha,ij} 
      - \overline{p}_\alpha \delta_{ij} $,
 and 
 $ \widetilde{D}_{\alpha,ij} 
    = 
    (  \partial_i \widetilde{u}_{\alpha,j} 
     + \partial_j \widetilde{u}_{\alpha,i}) /2 
    -  
    \left( \nabla \cdot \widetilde{\vct{u}}_\alpha \right) 
        \delta_{ij}/3.$
    Here $-\overline{p}\nabla \cdot \widetilde{\vct{u}}$ and $-\overline{\Pi}\mathrm{:}\widetilde{\vct{D}}$ denote the isotropic and anisotropic contributions to the pressure-strain interaction,
respectively
\citep[see, e.g.,][]{yang2022pressure}.

If one now takes the ensemble average and assumes that the dynamic viscosity $\mu$ is constant and independent 
of $\Sell$ for a given lag, then $\mu$ at any given scale $\ell$ can be calculated as
\begin{equation}\label{eqn:viscosity}
    \mu(\ell) 
        = 
    \frac{\langle \Phi^{uT}\rangle}
         {2\langle \Sell \mathrm{:} \Sell\rangle}
    = 
    \frac{\langle -\overline{p}\nabla \cdot \widetilde{\vct{u}} \rangle - \langle \overline{{\Pi}} \mathrm{:} \widetilde{\vct{D}} \rangle}
         {2\langle \Sell \mathrm{:} \Sell\rangle}.
\end{equation}
Finally, the kinematic viscosity can be calculated as $\nu(\ell)=\mu(\ell)/\rho$.

There is a slight complication associated with the above analogy since the VM model is compressible and it is being compared with incompressible MHD. 
Previous studies~\citep{yang2024effective} suggest that pressure dilatation 
    $-\overline{p}\nabla \cdot \widetilde{\vct{u}} $ 
is not necessarily related to viscous dissipation, but in the present 
termwise analogy
approach 
it is not clear how $-\overline{p}\nabla \cdot \widetilde{\vct{u}} $ can be related to the other terms in Eq.~\eqref{eqn:KE_MHD}. In fact, the viscosity estimate provided here accounts for both incompressible and compressible dynamical effects, while an effective viscosity estimated without using $-\overline{p}\nabla \cdot \widetilde{\vct{u}} $ cannot account for all compressible dissipation effects. 
A recent study of the Helmholtz decomposition~\citep{adhikari2025revisiting} of the pressure-strain interaction has shown that $\text{Pi-D}$ ($-\overline{{\Pi}} \mathrm{:} \widetilde{\vct{D}}$) contains a compressive element within it,
which implies that the incompressible viscosity should be estimated using only the incompressible element and not the full
$\text{Pi-D}$.
This refinement will be deferred to a later study.

A similar analogy 
is expected to hold for the magnetic energy evolution equations in MHD and VM systems, although investigation of this is also
deferred to a later study.
Note that Eq.~\eqref{eqn:KE_MHD} is not valid for electrons and so one cannot use the analogy to estimate an electron viscosity. 
Instead, in  \S\ref{sec:e-visc} 
we will make use of a different approach, one based on theory,  to estimate an effective viscosity for electrons.

In what follows, our approach is
to employ 
scale-filtering on data obtained from particle-in-cell (PIC) simulations of decaying turbulence 
to estimate an effective ion viscosity. We then perform MHD and two-fluid simulations---with the 
viscosity set to the value estimated from the PIC simulation---and compare the dynamics between these three types of simulations. 
 
The remainder of the paper is organized as follows: In section~\ref{sec:sim} we discuss the details of the simulations, 
with section~\ref{sec:results} presenting estimates for the effective viscosity and analysis of the results. 
Section~\ref{sec:conclu} provides the conclusions and implications of this study.

    \section{Simulations}\label{sec:sim}

We analyze three different types of freely decaying (unforced) simulations. 
First, we perform fully kinetic PIC simulations of decaying turbulence using the 
P3D code \citep{zeiler2002three}, and use the outputs to estimate the effective viscosity via Eq.~\eqref{eqn:viscosity}.
Second, we perform 
 two-fluid (electron inertial Hall-MHD: EIHMHD) 
   \citep{andres2014tf}
and MHD
simulations  with initial conditions set to be very similar to those of the main PIC run, and resistivity and viscosity equal to the (effective) viscosity estimate obtained from the PIC simulation. 

All simulations are performed in a $2.5$D setup 
with turbulent fluctuations in the $X\text{-}Y$ plane but no spatial variation in the $Z$-direction. 
Results are presented in normalized P3D units where lengths are normalized to the ion inertial length $d_i=c/\omega_{pi}$, 
with $c$ the speed of light 
and $\omega_{pi}$ the ion plasma frequency. 
Time is normalized to the 
inverse of the proton cyclotron frequency, $1 / \omega_{ci} $, 
and speeds to the Alfv\'en speed: 
   $ V_A = {B_0}/{\sqrt{4\pi n_0 m_i}}$ calculated using $B_0$, the arbitrary normalizing magnetic field 
and $n_0$ the normalizing density.

For the PIC simulations
the domain is a periodic square of side 
  $ L_\mathrm{box} \simeq 150\,d_i$ 
  with
  $4096$ grid points along each Cartesian axis.
For each species there are
  $3200$ particles per grid cell (PPG), 
  yielding about $107$ billion total particles.
The grid spacing is the electron Debye length $ dx = \lambda_{D_e}$, 
the mass ratio is $m_i/m_e=25$, 
background density $n = 1n_0$, 
and the plasma beta for both species is the same: $\beta_i=\beta_e=0.6$. In these units the wavenumbers $k$ corresponding to the proton inertial ($k_{d_i}$), electron inertial ($k_{d_e}$) and Debye  ($k_{\lambda_D}$) scales are located at $23.8$, $120$, and $628 \, k_\mathrm{box}$, respectively, with  $ k_\mathrm{box} = 2 \pi / L_\mathrm{box} $.
Initially, the root mean square (rms) fluctuation in velocity and magnetic field are equal: 
 $ \delta b_\text{rms} = \delta v_\text{rms}$. 
Initial conditions 
are created using random phased fluctuations for those Fourier modes whose wavenumbers  
satisfy $ 2 k_\text{box} \leq k \leq 4 k_\mathrm{box}$. 
The system is evolved without external forcing for more than 
    $ 10 \, \tau_\mathrm{nl}$, 
where $ \tau_\mathrm{nl}$ 
is the non-linear time estimated as 
  $ \tau_\mathrm{nl} = L_\mathrm{box} / \delta Z$, 
with 
  $ \delta Z = 
    \sqrt{  (\delta v_\mathrm{rms})^2 
          + (\delta b_\mathrm{rms})^2} $
the turbulence amplitude.
Such initializations 
are typical of the Alfv\'enic initial conditions used in decaying simulations of MHD turbulence~\citep{bandyopadhyay2018finite}. 

In addition to this main PIC simulation we perform several others, where either the PPG count is reduced
or the initial ion plasma beta is varied.
These provide information regarding scalings and numerical convergence.

For the fluid MHD simulations, both types, 
the domain is a periodic square box of side $L= 2\pi L_0$, 
where $L_0$ is a characteristic length. 
To facilitate comparison of the fluid and VM models we choose  $L_0$
to correspond to the correlation scale ($\approx$ the energy-containing scale) 
of the initial fluctuations, 
themselves chosen to be equivalent to the initial $\vct{u}$ and $\vct{b}$ in the PIC simulation.
Thus,  
 $ L_0 ={\int\left[ E(k)/k \right] \d k} / {\int E(k) \, \d k}$, 
where $E(k)$ is the spectral energy density at wavenumber $k$. 
The aliasing effects are suppressed by imposing 
  $ k_\mathrm{max} = N/3 $
as the maximum nonzero wavenumber. 
Here $N=4096$ is the total number of grid points along each Cartesian direction. 

A notable feature of this study is that the 
MHD simulations are initiated
with equal viscosity and resistivity, the values
are based on the estimation given in Eq. (\ref{eqn:viscosity})
from the PIC simulation.
The MHD/fluid runs adopt a 
value of total plasma $\beta= 1$. 
For the EIHMHD run, we use the same mass ratio employed in the PIC runs ($ m_i / m_e = 25 $), so that $ k_{d_i} \simeq 25 $, and $ k_{d_e} \simeq 125 $.
Run parameters 
are such that the dissipation wavenumber, 
 $ k_\mathrm{diss}(t) = \avg{ j^2 + \omega^2 }^{1/4} /\sqrt{\nu} $, 
 {where $\nu$ is the kinematic viscosity,}
has a maximum that is larger
than $ k_{d_e}$; 
specifically, 
 $ \max_t{ \left\{ k_\mathrm{diss}(t) \right\} } \sim 154$. 
This ensures that the EIHMHD run resolves the electron dynamics.



    \section{Results\label{sec:results}}

    \subsection{Effective ion viscosity}
         \label{sec:effect-ion-visc}

\begin{figure*}  
\centering
\includegraphics[width=\textwidth]{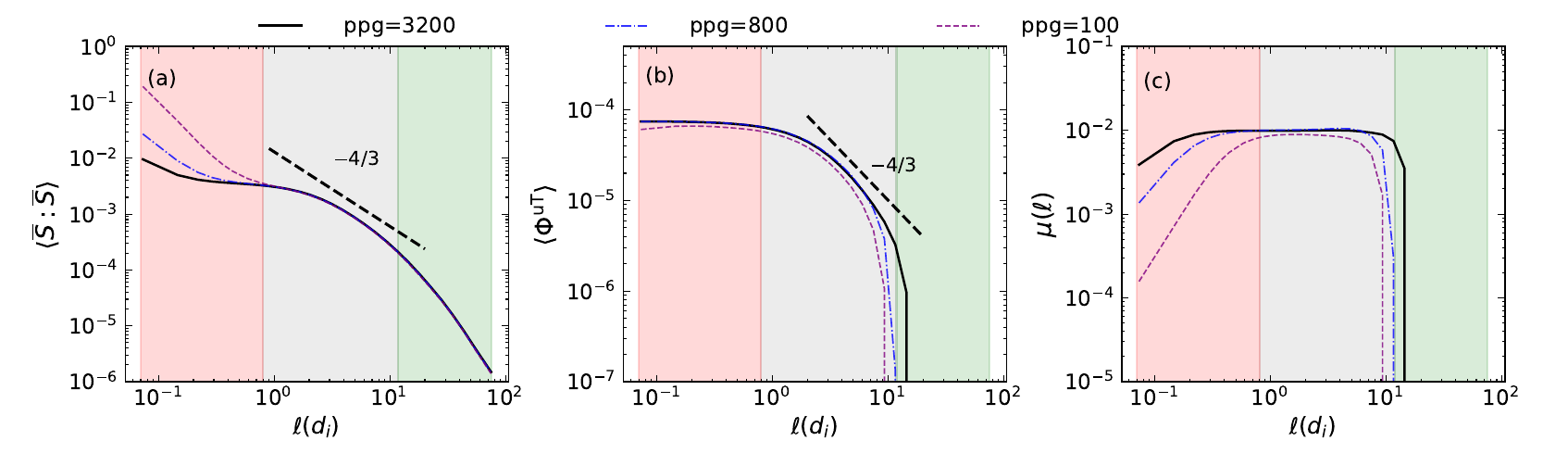}
  \caption{\label{fig:viscosity_estimate}Scaling of
  (a) Squared and traced scale-filtered rate-of-strain tensor,
  (b) filtered pressure-strain interaction,
  and (c) estimate of ion viscosity, 
  all as obtained from PIC simulations. 
  Results from three simulations with different values for the number of particles per grid cell (PPG) are shown.
  An $\ell^{-4/3} $ line (black dashed) is drawn for reference.}
\end{figure*}
In Fig.~\ref{fig:viscosity_estimate}, 
we show the trace of the squared rate-of-strain tensor 
$\langle \Sell \mathrm{:} \Sell \rangle$, 
the filtered pressure-strain interaction 
$\langle \Phi^{uT} \rangle$, 
and the effective viscosity $\mu(\ell)$, 
each as a function of lag $\ell$.
Here the angle brackets indicate spatial averaging rather than ensemble averaging.
Data is obtained from PIC simulations at $t\omega_{ci}\approx 3.5 \, \tau_\mathrm{nl} $, 
when the system is in an
approximately
statistically-steady state. 

In Fig.~\ref{fig:viscosity_estimate} we 
divide the lags into three different regions: the energy-containing, inertial, and kinetic ranges; in the figure these are denoted using green, gray, and red background-shaded rectangles, respectively. 
It is immediately evident from Fig.~\ref{fig:viscosity_estimate} that both 
$\langle \Sell \mathrm{:} \Sell \rangle$ 
and $\langle \Phi^{uT} \rangle$
are larger in the kinetic range and eventually fall 
off in the energy-containing range. 
Moving toward the smaller kinetic range of scales $<d_i$,
we note that 
$\langle \Phi^{uT} \rangle$
stays constant throughout the kinetic range, 
while $\langle \Sell \mathrm{:} \Sell \rangle$ 
increases as one approaches smaller scales $\ell \leq 0.2 \, d_i$. 
To examine this increase we perform additional simulations with reduced numbers of PPG: 800 and 100 vs.\ 3200. 
We find that $\langle \Sell \mathrm{:} \Sell \rangle$ is sensitive to the number of PPG and is 
inherently affected by the noise associated
with the discrete PPG
effect.\footnote{Recall that the Vlasov equation formally emerged in the limit of an infinite number of particles per Debye sphere.}
This is clearly an artifact of PIC simulations. 
The data suggests that the viscosity $\mu(\ell)$ would become constant at 
very small $\ell$ in the limit of infinite PPG. 


The effective viscosity $\mu( \ell)$,  
  Fig.~\ref{fig:viscosity_estimate}c, 
is found to be constant in the inertial range and part of the kinetic range. The constancy of the effective viscosity in the inertial range can be understood in terms of power scaling. 
The squared and traced
rate-of-strain tensor scales as $(\delta u/l)^2$. 
Following Kolmogorov-like turbulent energy spectrum, the velocity fluctuations $\delta u$ scale as $\ell^{1/3}$ resulting in $ \Sell \sim \ell^{-4/3}$. 
However, the scaling of the pressure-strain interaction is not straightforward to determine from its composition. Since the pressure-strain interaction is the contraction of the pressure tensor with the rate-of-strain tensor, the Fourier transform of the pressure-strain interaction is the convolution of the Fourier transforms of the pressure tensor and the rate-of-strain tensor. As a result, the theoretical scaling of the pressure-strain interaction is difficult to estimate. 
Empirically, the PIC simulation results seen in Fig.~\ref{fig:viscosity_estimate}b, 
suggest that the pressure-strain interaction exhibits a $\ell^{-4/3}$ dependence. As a result, the effective viscosity, 
which is the ratio of the filtered rate-of-strain tensor and filtered pressure-strain interaction, 
is independent of lag at inertial range scales and also into the kinetic range, displaying an extended plateau at 
    $ 
     \mu( \ell) = 0.01$.
We denote this plateau value as $ \mu^\mathrm{eff}$ and use it as
the estimated value of effective viscosity for ions. 
At smaller lags, $\mu( \ell) $ falls off, probably due to the particle noise effect, since 
 $ \langle \Sell \mathrm{:} \Sell \rangle$ 
increases at these scales. 
In addition, the maximum lag for the constancy of the effective viscosity also decreases with reduced PPG. 
Once again we emphasize that in the limit of an infinitely large number of particles per cell, 
we expect the effective viscosity to remain constant over a wide range of lag scale and extend all the way to the smallest lag in the system.

In the energy-containing range, the magnitudes of both $\langle \Sell \mathrm{:} \Sell \rangle$ 
and $\langle \Phi^{uT} \rangle$
are significantly reduced. 
Moreover, since 
  $ \langle \Sell \mathrm{:} \Sell \rangle$ 
is positive definite, 
but
  $\langle \Phi^{uT} \rangle$
is not, 
using Eq.~\eqref{eqn:viscosity} to determine the effective viscosity means it could be negative for some scales, 
  with this being more likely in the energy-containing range.
However, since the net viscous dissipation in the energy-containing range is expected to be negligible, 
there is little need to employ an effective viscosity in this range. 

Using the effective viscosity and mean square vorticity $\avg{ \omega^2} $, 
one may calculate the rate of 
effective viscous\footnote{Here we are neglecting compressive contributions to viscous dissipation, which, in a fluid model, would be 
 $ \propto \zeta \avg{ (\nabla \cdot \vct{u})^2}$,
 where $\zeta$ is the bulk viscosity.} 
dissipation as 
  $ D_\mu 
     = \mu^\mathrm{eff} \avg{ \omega^2 } $ 
and compare it with the rate of change of ion flow energy $ - \partial \avg{E_f}/\partial t$. 
At the time of analysis, we find  
 $ D_\mu = 1.02 \times 10^{-4}$ 
and 
 $ - \partial \avg{E_f} / \partial t = 9.19 \times 10^{-5}$.
The approximate equivalence of these values provides a validation of the effective viscosity approach.

The effective viscosity estimated above is the dynamic viscosity of ions. We may also calculate an effective kinematic viscosity, 
  $ \nu^\mathrm{eff} = \mu^\mathrm{eff} / \langle \rho\rangle =0.01$, 
where 
 $ \langle \rho \rangle  = m_i \langle n_i\rangle = 1 $ 
is the mass density since both the mass and the ion density are unity in the setup for this run.


We may compare these results to those obtained via an alternative (non scale-filtering) approach for estimating an effective viscosity~\citep{yang2024effective}.
In that approach the effective viscosity is not determined as a function of scale so that
the estimate would probably account for the viscosity at the smallest scale in the simulation. 
Adopting this correspondence, 
the effective viscosity estimate $\mu^\mathrm{eff}_{\ell\sim 0.06}=3.9\times 10^{-3}$ 
  (see Fig.~\ref{fig:viscosity_estimate}c)
is close to the estimate obtained by~\citet{yang2024effective}, 
 $ \mu = 5.31 \times 10^{-3}$.

   \subsection{Kolmogorov scales and Reynolds number}
   \label{sec:KolScales}
   
Having accomplished an estimation of effective viscosity for nearly collisionless plasma turbulence, 
we can now calculate the (effective) Kolmogorov scales for length $\eta$, time $\tau_\eta$, and velocity $u_\eta$
  \citep{Kol41a,Frisch}. 
For an incompressible fluid these scales, by definition, depend only on the (spatially-averaged)
total energy dissipation rate (per mass) $\epsilon$ 
and the kinematic viscosity $ \nu $:
\begin{equation}\label{eqn:kolmogorov_scale}
    \eta = \left(\frac{\nu^3}{\epsilon}\right)^{1/4}, \quad
    \tau_\eta = \left( \frac{\nu}{\epsilon}\right)^{1/2}, \quad
    u_\eta = \left( \nu \epsilon \right)^{1/4}.
\end{equation}


For the kinetic system,
the energy dissipation rate can be calculated using the time rate of change of the volume-averaged
magnetic $\avg{E_B}$ and 
ion flow $\avg{E_{if}}$ energies,\footnote{The 
  rate of change of electron flow energy is negligible relative to the net dissipation in the system~\citep{yang2024electron}.} 
i.e.,
\begin{equation}
     \avg{\rho} \epsilon 
   = 
   - \partial \avg{E_B + E_{if}} / \partial t
        \doteq 5.31 \times 10^{-4}.
\end{equation}
Here we have employed the approximation, akin to Favre averaging, that 
  $\avg{ \rho(\vct{x}) \epsilon(\vct{x})} \approx \avg{\rho} \epsilon $ on the LHS. Employing this estimate for $\epsilon$ in Eq.~\eqref{eqn:kolmogorov_scale}, 
and setting $ \nu = \nu^\mathrm{eff} $,
yields
    $      \eta = 0.20 $, 
    $ \tau_\eta = 4.34 $, 
and $    u_\eta = 0.048 $. 
The Reynolds number based on the large-scale flow features can be computed using
\begin{equation}
    Re = \left(\frac{L}{\eta}\right)^{4/3} = 135,  
\end{equation}
where $L$ is the correlation length estimated using an average value of the initially excited Fourier wavenumbers, 
{\it viz.}     
  $ L = 1 / k_\mathrm{av} = L_\mathrm{box} / (3\times 2\pi) = 7.93$.
This is consistent with the approximately one decade inertial range present in this system 
  (see ~\citet{parashar2018dependence, adhikari2021energy}). 
We remark that this estimated Reynolds number is smaller than the estimate provided in~\citet{yang2024effective}, 
primarily because the effective viscosity estimated in this paper is approximately twice the estimate of \citet{yang2024effective}.

   \subsection{Ion viscosity and temperature}
   \label{sec:ion-visc-temp}
   
\begin{figure}
\centering
  \includegraphics[width=1\linewidth]{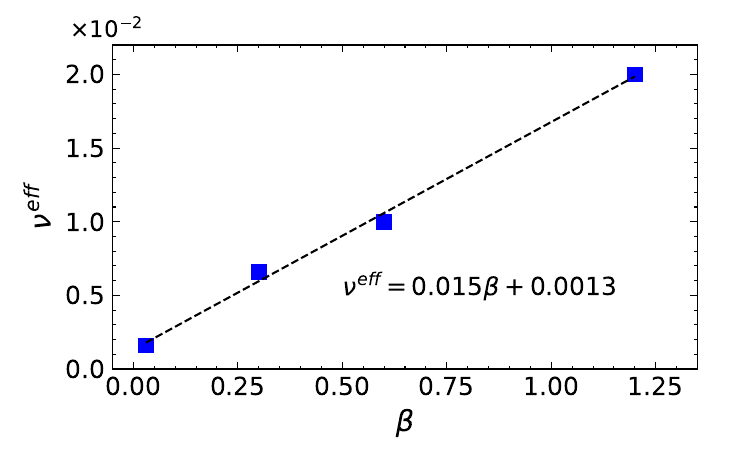}
  \caption{\label{fig:viscosity_beta}Effective viscosity of ions as a function of ion plasma beta. The dashed line is the line of best-fit.}
\end{figure}

Next we explore the effect of plasma beta $\beta$ on the effective viscosity of ions,
by analyzing three additional runs 
with the same initial conditions but different ion $\beta$ values, namely $0.03$, $0.3$ and $1.2$. 
The value of $\beta$ is adjusted by changing the (initial) ion temperature. 
As is seen in Fig.~\ref{fig:viscosity_beta}, 
we find that the (plateau) effective viscosity, 
  $ \nu^\mathrm{eff}$,
increases with increasing ion plasma $\beta$ (and thus, with increasing ion temperature).  
While previous studies have shown an inverse dependence of viscosity on temperature in dusty plasmas~\citep{haralson2016temperature}, the present finding is consistent with the effect of temperature on the viscosity of neutral gases. 

   \subsection{Effective electron viscosity}
   \label{sec:e-visc}

Although it is invalid to use 
    Eq.~\eqref{eqn:viscosity} 
directly for electrons, we may still estimate an effective electron viscosity by exploiting a relationship between it and the ion viscosity.
Following~\citet{braginskii1965reviews}, there has been some significant progress in estimating the viscosities for ions and electrons in fully ionized plasmas for all values of the ion charge $Z$~\citep{whitney1999momentum}. 
For a collisional electron-proton
plasma the ratio of ion and electron viscosities behaves as~\citep{velikovich2001role}
\begin{equation}
    \frac{\mu_i}{\mu_e}=\left(\frac{T_i}{T_e}\right)^{5/2} \left( \frac{m_i}{m_e}\right)^{1/2}.
 \label{eq:mu-i-mu-e}
\end{equation}

For our PIC simulations 
  $ m_i/m_e=25 $ 
and 
 $ T_i/T_e = 1 $, 
so that 
  $ \mu_e = \mu_i / 5 = 2\times 10^{-3} $,
 where we assume that the formula also holds for weakly collisional plasmas.
Magnetospheric Multiscale Mission (MMS) observations have typical temperature ratios of 
    $ T _i / T_e = 5 $, 
and the true physical mass ratio  
    $ m_i / m_e = 1836 $, 
yielding 
 $ \mu_i \simeq 2400 \, \mu_e $, 
implying that the ion viscosity is much larger than the electron viscosity.\\
Note that 
\citet{yang2024effective} using MMS data
found a different result, namely that $\mu_i/\mu_e \simeq 150$.
This discrepancy may be due to inaccuracy of the approximation
in Eq. \ref{eq:mu-i-mu-e} for plasmas of low collisionality such as the magnetosheath.


    \subsection{Comparison with Two-fluid and MHD simulation}
    \label{sec:2fluid-mhd-comp}
    
In this subsection, we compare the energetics of our main kinetic (PIC) simulation with results from 
the two-fluid (EIHMHD) and MHD simulations. 
Recall that these are performed with initial conditions very similar to those of the PIC run and using a viscosity equal to the
  (inertial range plateau)
effective viscosity estimate obtained from the PIC simulation. 
Note, however, that because the PIC and fluid codes employ different normalizations and domain sizes,
 $ \nu^\mathrm{eff}_{\text{PIC}} = 0.01 \, V_A d_i $
corresponds to 
  $ \nu^\mathrm{eff}_{\text{fluid}} = 0.01 \times (2\pi/L_\text{box}) \, V_A L_0 
 $
for both the fluid cases,
where $L_\text{box}$ is the length of the box for the PIC simulation.

\begin{figure}
\centering
\includegraphics[width=\columnwidth]{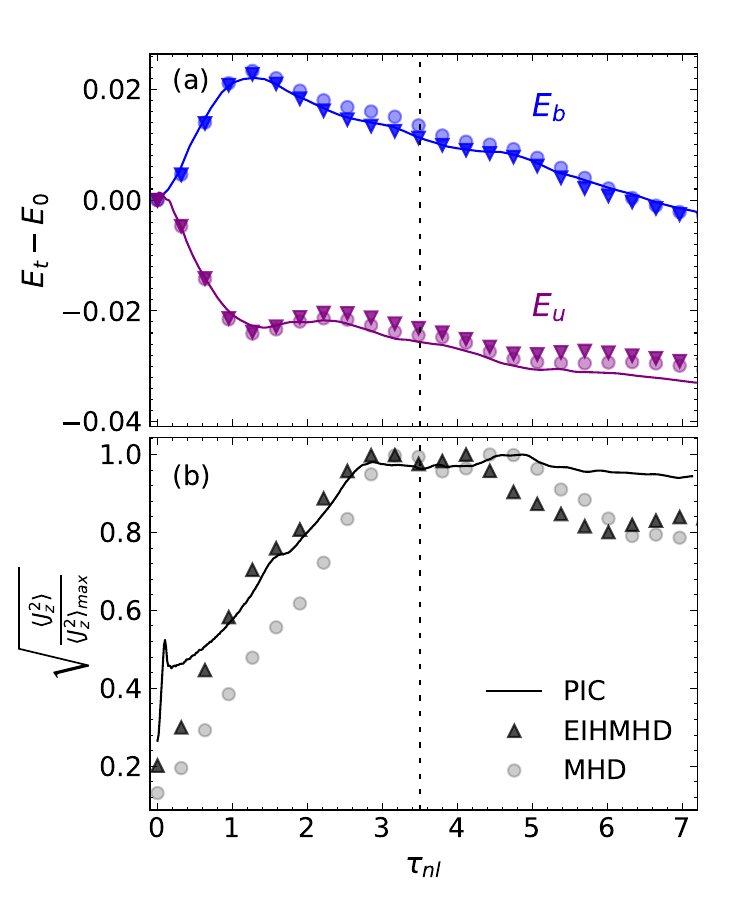}
\caption{\label{fig:energy_comp}Time evolution 
 of some equivalent quantities from the kinetic, two-fluid, and MHD simulations:
 (a) the change in spatially-averaged 
 magnetic energy $E_b$ and 
 ion flow energy $E_u$,
 and
 (b) the normalized rms  out-of-plane electric current density, $j_z$.
 Normalization of the rms $j_z$ is with respect to its maximum in the time series. 
 Note that the viscosity in the two-fluid and MHD simulations is equivalent to the effective viscosity obtained from the PIC simulation.
 The vertical dashed line denotes the time of analysis used in connection with Fig.~\ref{fig:viscosity_estimate}.}
\end{figure}

In Fig.~\ref{fig:energy_comp}a we compare the time evolution of the magnetic energy and ion flow energies for these three cases. The change in magnetic energy in the two-fluid case overlaps almost perfectly with the PIC results, while the magnetic energy in MHD has a 
slightly ($\sim 10\% $) larger value during the period 
  2--4\,$ \tau_\mathrm{nl} $. 
Turning to the
  (change in) the ion flow energy for these runs,
these also follow each other very closely until about 
 $ 2 \, \tau_\mathrm{nl}$.
At later times, the two-fluid and MHD simulations both show smaller decreases ($\approx 10\%$) in ion-flow energy compared to the kinetic case.
The slight discrepancy towards the end is possibly due to the differences in channels of energy transfer and dissipative mechanisms in these systems. For example, the pressure-strain interaction, a pathway that relates flow energy to thermal energy is absent in both the MHD and EIHMHD cases.

Fig.~\ref{fig:energy_comp}b displays the time evolution of the rms value of the (normalized) out-of-plane electric current density, $j_z$, across these three simulations.
Here too, 
the evolutions are rather similar, particularly in the period where $j_{z,\mathrm{rms}} $ is approximately steady ($ \tau_\mathrm{nl} \sim 3$--$4$),
corresponding to well-developed turbulence. 
During the early and late phases of the simulations, 
$j_{z,\mathrm{rms}}$
levels for both of the fluid runs fall below the level from the kinetic simulation. This is because in the PIC case, the initial magnetic islands and subsequent secondary islands can interact at the electron scales resulting in sharp gradient of the magnetic fields and therefore larger out of plane currents.

\begin{figure}
\centering
\includegraphics[width=\columnwidth]{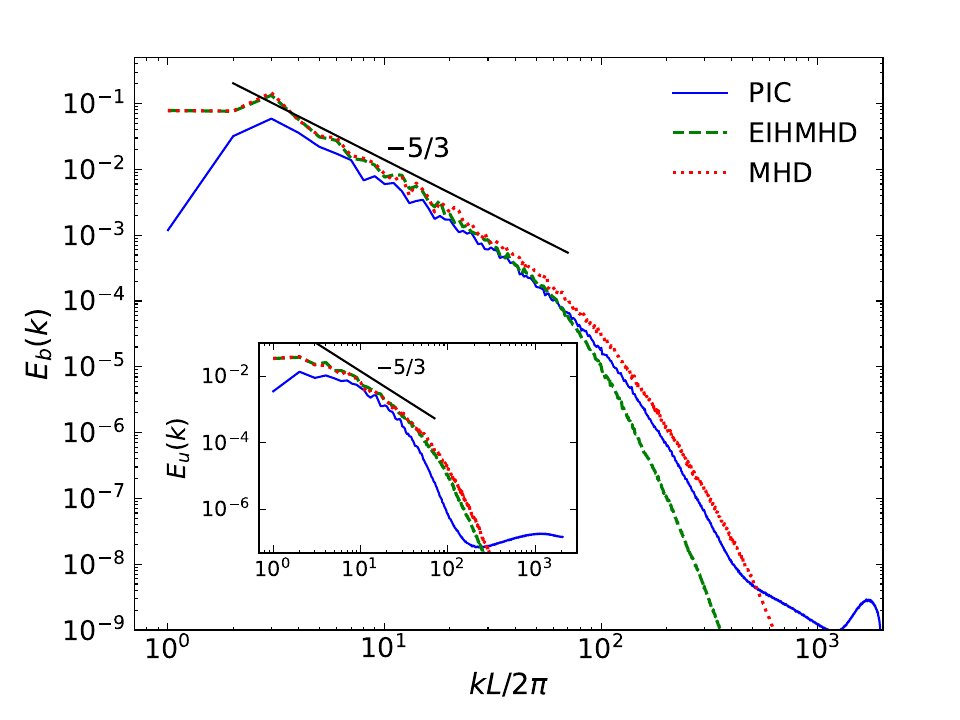}
\caption{\label{fig:spectra_comp} 
A comparison of the magnetic energy spectra for MHD, EIHMHD and PIC simulations at the time of analysis $t=3.5 \tau_{nl}$, shown by the dashed line in Fig.~\ref{fig:energy_comp}. In the inset we compare the kinetic energy spectra. The solid reference lines correspond to a slope of $-5/3$.}
\end{figure}

In Fig.~\ref{fig:spectra_comp}, we compare the magnetic and kinetic energy spectra for these three cases at the time of analysis $t=3.5\tau_{nl}$. Clearly, the magnetic energy spectra are almost identical at the intermediate wavenumbers $3\leq k \leq 50$, following a $-5/3$ Kolmogorov-like spectrum. This suggests that these systems, initialized with identical conditions, follow similar time evolution in the inertial range. While the magnetic spectra for PIC follows MHD closely at higher wavenumbers, the spectra for EIHMHD deviates from the MHD and PIC and exhibits a relatively steeper slope. At large wavenumbers, we see a small peak for the magnetic energy spectrum for PIC, which is often characterized to the discrete particle effect (noise). The inset of Fig.~\ref{fig:spectra_comp} shows that the kinetic energy spectra of these three cases follow each other closely for wavenumbers $3\leq k \leq 50$ and exhibit a $-5/3$ slope. However, for $k>30$, the kinetic energy spectrum for the kinetic case, falls off earlier than the other two cases. While the inertial scale features of both the kinetic and magnetic spectra are similar in these system, there exist some differences in the smaller (large length scale) and larger wavenumbers (smaller length scales). This is mostly likely due to the differences in energy transfer mechanisms at these scales.

    \section{Conclusion and Discussion}\label{sec:conclu}

In this paper, using kinetic particle-in-cell simulations of nearly collisionless plasma turbulence, we estimate the effective viscosity. We use a scale-filtering (or coarse-graining) approach to compare the time evolution equation of the flow energy in both 
magnetohydrodynamic (MHD) and Vlasov--Maxwell (VM)
formalisms. We find that the effective viscosity, while formally  scale dependent, become
independent of scale in the inertial range and even into the kinetic range. 
We hypothesize that the decrease in the value of the effective viscosity seen at the smallest scales is solely due to the limited number of particles per grid cell used in our PIC simulations. Therefore our conjecture is that for arbitrarily large numbers of particles per cell, 
the derived effective viscosity would become constant down to arbitrarily small lags.
At the larger (e.g., energy-containing range) scales, effective viscosity is insignificant, since the net dissipation at those scales is negligible. 

There are also interesting differences in runs with different PPG in the large scale end of the inertial range and into the 
energy containing range. While almost no difference with variation in PPG is seen in the scale filtered trace of the rate-of-strain at these outer scales,
there are quite noticeable deviations in the PPG dependence of the pressure-strain and the effective viscosity as scales approach the correlation length. 
But the effective viscosity becomes negligible at these scales so the disparity with 
varying PPG becomes of less importance. 

The present estimate of effective viscosity is based on the full pressure-strain energy conversion channel, including the 
pressure-dilatation which is manifestly compressive in nature. 
The estimation of viscosity given by \citet{yang2024effective} made use of a different method, 
conditioning $\text{Pi-D}$ on the traceless rate-of-strain tensor $D$. 
Therefore, the contribution due to pressure dilatation is present in 
our estimate but not in that of \citet{yang2024effective}. 
This at least partially accounts for the somewhat larger value of effective viscosity that we find, 
even if typically values of $p\theta$ are usually smaller that values of $\text{Pi-D}$ in standard 
turbulence simulations that lack large scale compression or expansion.

We also performed an MHD and a two-fluid simulation with initial conditions that are almost identical to those of the PIC run.
For these runs the imposed values of viscosity and resistivity were set equal to 
the effective viscosity obtained from the PIC simulation. 
Upon comparing the results of these three cases, we found striking similarities in the global behavior of magnetic and ion bulk flow energy, 
We also found close similar behaviors in the time history of the 
out-of-plane current densities, particularly near the time of maximum dissipation.
These consistent aspects of the three types of simulations 
provide validation of main result, which is the method for estimation of the effective viscosity.

In future work we expect to examine the assumption made in Eq. \ref{eqn:viscosity} 
that $\mu$ and $\Sell$ are statistically independent. This should be 
checked a posteriori using simulation data, even though the 
constancy of $\mu$ in the results suggest that the approximation is valid.

In computing characteristic turbulence scales based
on the effective viscosity we found that the 
Kolmogorov scale $\eta$
is smaller than the ion inertial length $d_i$.
This may seem to be an anomalous result given that $d_i$ is often associated with the spectral break at the 
termination of the inertial range in the solar wind at order one or high plasma $\beta$ \citep{leamon1998observational, chen2014ion}. 
The fact that here the estimated $\eta$ is smaller than $d_i$ seems decidedly not hydrodynamic-like.
There is however, precedent for this, an example being the 
study of Taylor scale $\lambda_T$ in the solar wind
~\citep{matthaeus2008interplanetary}.
Normally in hydrodynamics turbulence one expect that $\lambda_T > \eta$.
But~\citet{matthaeus2008interplanetary} found that solar wind with low plasma $\beta$ 
are characterized by estimated values of $\lambda_T < \d_i$. This kind of non-hydrodynamic scaling of 
turbulence parameters is perhaps expected in collisionless plasmas for which there are numerous additional
physical length scales present, including those associated to both electrons and ions.

In closing, the present 
approach for estimating viscosity
may have broad application in collisionless plasmas. 
One immediate extension is to use this technique in analysis 
of MMS observations. The MMS datasets provide high resolution measurements that resolve scales as small as electron 
length scales. This introduces the possibilities of computing directly the electron viscosity which in turn provides a contribution
to resistivity.

\begin{acknowledgments}
This research is partially supported by the MMS Theory, Modeling and Data Analysis team under NASA grant 80NSSC19K0565, 
by the NASA LWS program under grants 80NSSC20K0198 and 
80NSSC22K1020, 
and a subcontract from the New Mexico consortium 655-001, 
a NASA Heliophysics MMS-GI grant through a Princeton subcontract SUB0000517, and by the National Science Foundation
Solar Terrestrial Program grant 
AGS-2108834. 
Y.Y. is supported by 2024 Ralph E. Powe Junior Faculty Enhancement Award and the University of Delaware General University Research Program grant.
This research was also supported by the International Space Science Institute (ISSI) in Bern, through ISSI International Team projects \#556 (Cross-scale energy transfer in space plasmas) and \#23-588 (Unveiling energy conversion and dissipation in nonequilibrium space plasmas).
We would like to acknowledge high-performance computing support from Cheyenne (doi:10.5065/D6RX99HX) and Derecho (https://doi.org/10.5065/qx9a-pg09) provided by NCAR's Computational and Information Systems Laboratory, sponsored by the National Science Foundation.

\end{acknowledgments}

\bibliography{main}{}
\bibliographystyle{aasjournalv7}



\end{document}